\documentclass[reprint,superscriptaddress,floatfix,amsmath,amssymb,pra,aps,showpacs,twocolumn]{revtex4-1}
\usepackage{amsmath,amsfonts,amssymb,amsthm,graphics,graphicx,epsfig,bbm}
\usepackage[colorlinks=true,citecolor=blue,linkcolor=blue,urlcolor=blue]{hyperref}
\usepackage[usenames]{color}
\usepackage{graphicx}
\usepackage{subfigure}
\usepackage{amsmath}
\usepackage{epsfig}
\usepackage{dcolumn}
\usepackage{bm}
\usepackage{color}
\usepackage{times}
\usepackage{epstopdf}
\usepackage{amssymb}
\usepackage{amstext}
\usepackage{latexsym}
\usepackage{hyperref}
\usepackage{amsfonts}
\usepackage{psfrag}
\usepackage{soul,xcolor}
\usepackage[normalem]{ulem}
\usepackage{dsfont}
\usepackage{txfonts}

\begin{document}


\title{Teaching labs for blind students: equipment to measure the thermal expansion coefficient of a metal. A case of study}
\author{O. Negrete}
\email[O. Negrete]{\quad oscar.negrete@usm.cl}
\affiliation{Departamento de F\'isica, Universidad T\'ecnica Federico Santa Mar\'ia, Casilla 110-V, Valpara\'iso, Chile.} 
\affiliation{ Centro para el Desarrollo de la Nanociencia y la Nanotecnolog\'ia, CEDENNA, 8320000 Santiago, Chile.}

\author{A. Lisboa}
\email[A. Lisboa]{\quad alfredo.navarro@usm.cl}
\affiliation{Departamento de F\'isica, Universidad T\'ecnica Federico Santa Mar\'ia, Casilla 110-V, Valpara\'iso, Chile.}%

\author{F. J. Pe\~na}
\email[Francisco J. Pe\~na]{\quad francisco.penar@usm.cl}
\affiliation{Departamento de F\'isica, Universidad T\'ecnica Federico Santa Mar\'ia, Casilla 110-V, Valpara\'iso, Chile.}%

\author{C.O. Dib}
\affiliation{Departamento de F\'isica, Universidad T\'ecnica Federico Santa Mar\'ia, Casilla 110-V, Valpara\'iso, Chile.}%
\affiliation{Centro-Cient\'ifico-Tecnol\'ogico de Valpara\'iso, CCTVal, Valpara\'iso, Chile.}
\author{P. Vargas}
\affiliation{Departamento de F\'isica, Universidad T\'ecnica Federico Santa Mar\'ia, Casilla 110-V, Valpara\'iso, Chile.}%
 \affiliation{ Centro para el Desarrollo de la Nanociencia y la Nanotecnolog\'ia, CEDENNA, 8320000 Santiago, Chile.}

\date{\today}

\begin{abstract}
We design a Teaching laboratory experience for blind students, to measure the linear thermal expansion coefficient of an object. We use an open-source electronic prototyping platform to create interactive electronic objects, with the conversion of visual signals into acoustic signals that allow a blind student to participate at the same time as their classmates in the laboratory session. For the student it was the first time he managed to participate normally in a physics laboratory.

\end{abstract}

\maketitle

\section{\label{intro}Introduction}

Measuring the coefficient of thermal expansion of an object is one of the common experimental experiences for undergraduate engineering students in a course on Thermodynamics. As we know, changes in temperature induce change in the size of a material object. 

In our experimental setup, we are interested in determining the change in length of a straight aluminum tube as a function of temperature, keeping the pressure constant at atmospheric value. The thermal expansion of an object of length $L$ can be specified by the relationship between the fractional change in length $\Delta L/L$ and the temperature variation $\Delta T$. This relation is characterized by the linear expansion coefficient $\alpha$, which depends on the material (aluminum, in our case) but not on the size of the object:\cite{Fermi_Thermo,Callen} 

\begin{equation}
\frac{\Delta L}{L_{0}}=\alpha\  \Delta T,
\end{equation}

where $L_{0}$ corresponds to the natural length of the tube at room temperature. For a blind student, it is impossible to obtain measurements for this experiment (or many other) because the measuring instruments are usually of visual character. Therefore, a change in the measuring devices are necessary in order to include this student in the lab experience \cite{Holt,DeBuvitz}. A natural alternative to visual readings is to include sounds that relate to the value of the physical variable that we intend to measure. An open-source electronic prototyping platform (Arduino board) offers the possibility to use a variety of sensors and electronic components that can be used to improve the data acquisition and control of different experiments in physics    \cite{Marinho,Goncalves1,Atkin,KuvNob,Goncalves2,Toenders,DeBuvitz,Galeriu1,Galeriu2,Galeriu3}. 

On the other hand, curricular adaptations for blind students have always been a big challenge. Many Universities, for example, of the United States have failed in the equal access that must be given for homework material,  classes and software for these students \cite{Zou,Lewin,Scanlon}.  Consequently, making correct protocols to work with this type of students in physics laboratories is necessary, interesting and in order.

In our University, the construction of knowledge in social environments is fundamental, which is why collaborative learning (CL)  \cite{Millis} is part of our educational model \cite{WEB3}. The elements that are always present in collaborative learning are cooperation, responsibility, communication, teamwork, and self-evaluation. A properly trained group activates safe and stimulating environments for work. It is for these reasons that it was promoted the realization of an experience in which the members of the group acquired different responsibilities with the interdependence of procedures and experimental data, which promoted the discussion and the generation of adequate learning.

In this work, we present the successful case of a measurement of the linear expansion coefficient for aluminum, done by a blind engineering student at our Institute. Our intervention consisted in  the modification of the experimental instruments used for the length measurement. The readings of this equipment are usually done visually. We devised a conversion to audible signals, with an accuracy comparable to the resolution of the instrument.


The student, who was visually impaired from a very young age, was attending his second year in Computer Engineering and had passed his first two courses ``Introduction to Physics'' and ``Mechanics'', where appropriate modifications were done in the lectures for him to complete the courses. Among the modifications were the use of tactile graphics (such as in Ref.\cite{Holt}) and the use of the ``Accessibility" option in Microsoft Word\texttrademark \  \cite{software1} that transcript the text into sounds. The student also showed excellent handling with Microsoft Excel\texttrademark\  \cite{software1} software.

In Section II we describe the laboratory experiment with the regular instrumentation and our modification for visually impaired students. In Section III we show the results and performance of the experiment operated by the student. In Section IV  we state our conclusions.

\section{\label{expsetup} Experimental Setup} 

\subsection{The Regular Setup} 

\begin{figure}
    \centering
    \includegraphics[width=1.0\columnwidth]{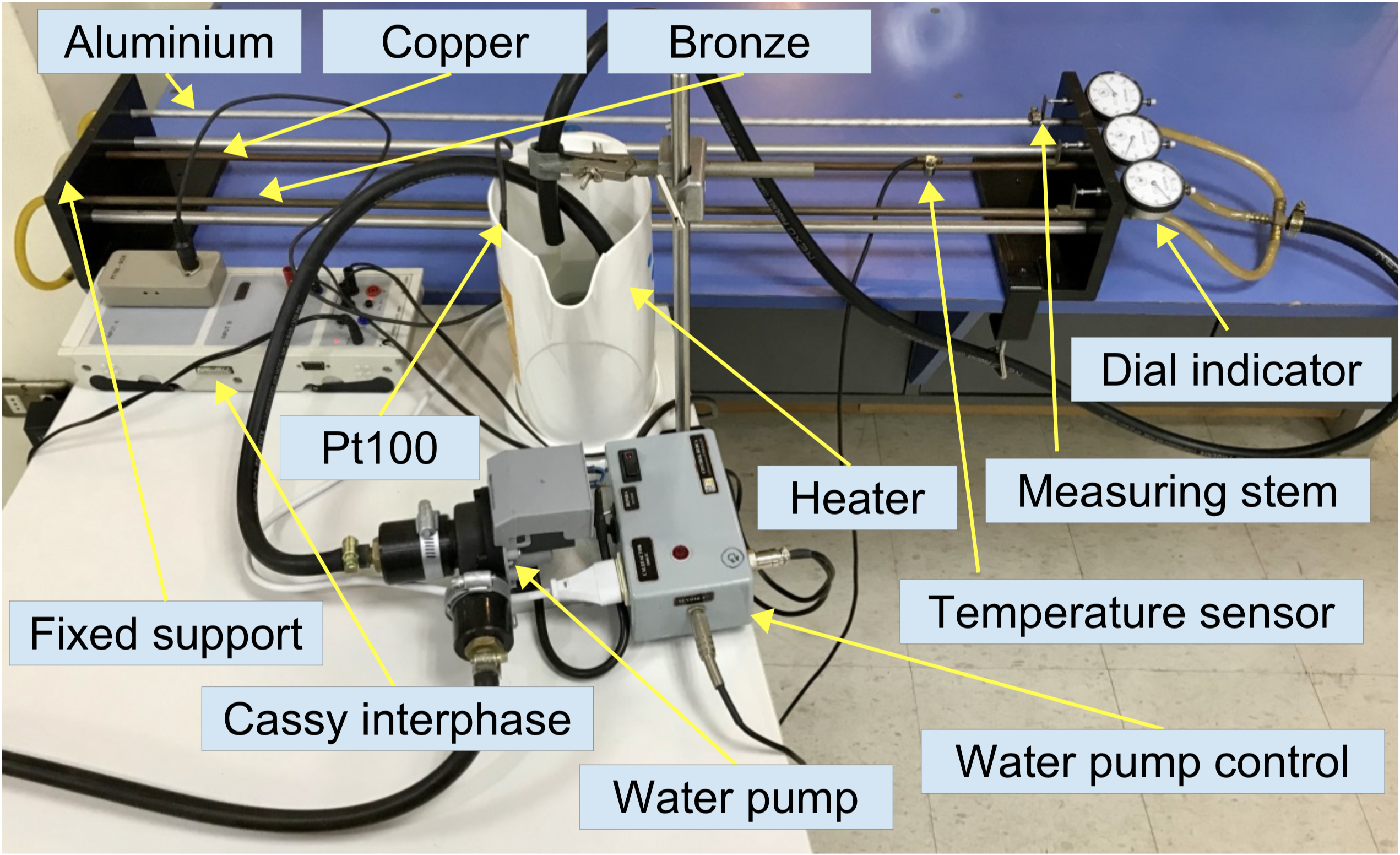}
    \caption{Standard system for measuring the linear expansion of metal tubes for regular students. The main parts are indicated. Three tubes to be measured are arranged top to bottom: aluminum-copper-bronze. The dial indicators where the visual reading is done are also indicated.  Students relate the displacement of the needle in these dial indicators with the longitude change of the corresponding tube. The Pt100 temperature sensor, the heater, the location of the temperature sensor, the water pump and the Cassy interface are also indicated.}
    \label{fig1}
\end{figure}

In this lab experience, we built a closed system (see Fig. \ref{fig1}) composed of a water pump whose function is to circulate the water,  an electric kettle responsible for increasing the temperature, and finally hoses connected to the tubes that bind the system. The temperature of the water is increased using a kettle power control that uses the temperature measured by the Pt100 as a reference. The Pt100 sensor is essentially an electric resistance that varies under changes in temperature. The fundamental feature of the Pt100 is that it allows temperature measurements in a linear scale because the platinum (material from which it is made) varies its electric resistance proportionality to the variation of temperature. 

The students have three tubes of different material each (aluminum, copper, and bronze), of 0.8 mm in diameter. The tubes are located on an insulating plastic base with one of their extremes fixed and the other free to allow their change in length.  The temperature of the tubes is measured with a solid-state sensor attached to the copper tube only (see Fig. \ref{fig1}) and the three tubes are considered to have the same temperature  ($T$), which is a good approximation given the expected precision of the experiment. The latter is the variable we need in the analysis of the thermal expansion coefficient.

They compare their results with the standard values given in \cite{WEB00,Freedman} for the linear thermal expansion coefficients:

\begin{equation}
\begin{split}
&\alpha_{Al}= 24.0 (21-24)\times 10^{-6} \ 1/{^\circ C}, 
\\ 
&\alpha_{Cu}= 16.6 (16.0-16.7)\times 10^{-6}  \ 1/{^\circ C}, 
\\ 
&\alpha_{Brz}= 18.0 (17.5-18.0)\times 10^{-6}  \ 1/{^\circ C},
\label{eq2}
\end{split}
\end{equation}

and determine the percentage error $(\xi)$ of their measurements as: 

\begin{equation}
\xi= 100 \times \left | \frac{ \alpha^{measured}-\alpha^{standard}}{\alpha^{standard}} \right |.
\end{equation}

It is important to remark that the standard values given above may not match precisely with the expansion coefficients of the materials used in our laboratory. This is simply because the standard values are given to materials with a specific micro-structure, while the materials in our laboratory do not have any specification other than their composition. However, in previous tests 
 the 
we have verified that the actual thermal expansion coefficients of the tubes in our lab coincide with the standard values of
Ref. \cite{WEB00,Freedman} within an error less than 0.2 $\%$.

Almost all of the steps in this lab experience can be performed today by a student with impaired vision, including those that require special software, except for the reading of the dial indicator. Specifically for this reading we designed special instrumentation that we present here and which is the subject of this article. The dial indicator is an enclosed instrument so that neither the needle nor the scale behind it are accessible to touch. 

Our purpose is to include a student with impaired vision in the measurement phase within the regular laboratory session together with his/her classmates. This laboratory experience is done in groups of two students.

In the actual lab session, we require the blind student to do all the measurements necessary to obtain the thermal coefficient of the aluminum tube only (i.e. $\alpha_{Al} $). 

In the next subsection, we describe the modifications that were made to carry out this experience, replacing the visual measurement of the dial indicator by a measurement of auditory type using an Arduino platform \cite{WEB0}.  

\subsection{Our Modified Setup}

\begin{figure}
    \centering
    \includegraphics[width=0.8\columnwidth]{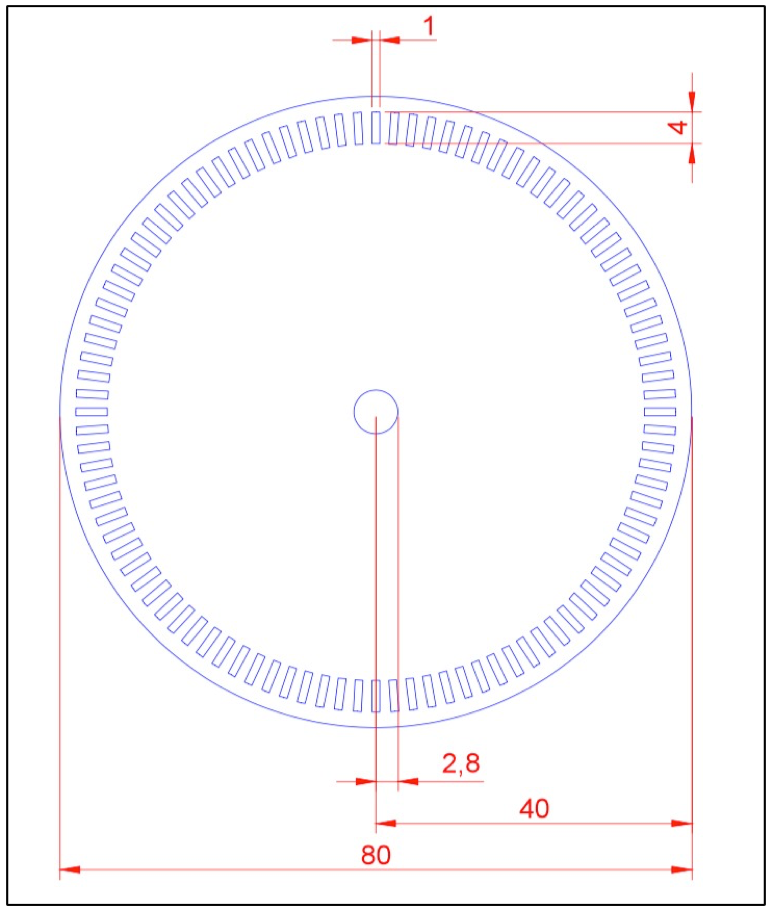}
    \caption{Modified version of dial indicator. The numerical values in this figure are given in mm.}
    \label{fig3}
\end{figure}

\begin{figure}
    \centering
    \includegraphics[width= 0.8\columnwidth]{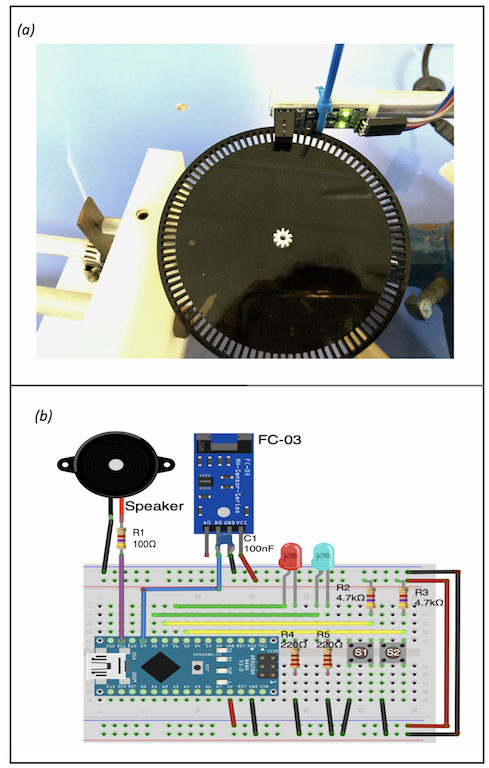}
    \caption{(a) Photogate FC-03 mounted with the modified disk made of black acrylic non-translucent. The disk has a mass of 13.3 gr. (b) 
A graphic representation of the experimental setup on a protoboard. Each electronic component is marked with their respective values along with the connection to the main board used in the final device version.
}
    \label{fig4}
\end{figure}

Our modification consists in replacing the dial indicator with a disk of 2.8 mm inner radius and 40 mm outer radius and 3 mm thick, with 100 equally spaced slits  $1$ mm $\times$ $4$ mm on the outer edge (see Fig. \ref{fig3}). The turn of the disk between consecutive slits corresponds to a 0.01 mm dilatation of the tube. The disk is made of black non-translucent acrylic with a mass of 13.3 gr. One point we considered in this design is that the disk should be large enough to have a good resolution with reasonably cheap optical devices (miniature devices are more expensive) and it should be small and light enough to have a low inertia for the mechanical parts. 

\begin{figure}
    \centering
    \includegraphics[width= 0.8\columnwidth]{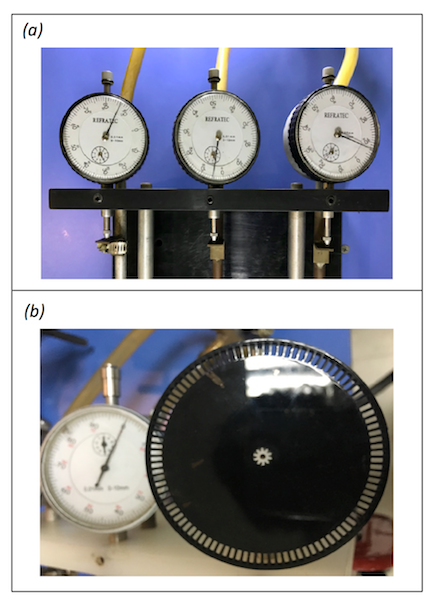}
    \caption{(a) Regular dial indicator. The long pointer have a resolution of 0.01 mm and the short pointer of 1 mm. The tubes from left to right; aluminum, copper and bronze. (b) Modified version of dial indicator that makes it compatible with a photogate. Every empty space corresponds to a measurement of 0.01 mm. One loop corresponds to 100 empty spaces giving a total per loop of 1 mm. The internal clock will not be used for this proposed experience due to the small changes in $\Delta L$ that our students register.}
    \label{fig2}
\end{figure}

The reading of the slits is done with a photogate sensor (model FC-03) generally used in small motor speed controls. The photogate sensor has a gap of 5 mm where the 3 mm thick disk can fit through  (see Fig. \ref{fig4}(a)). The dial indicator is replaced by the disk. This is done by removing the dial indicator from its axis, and mounting the disk in its place (see Fig. \ref{fig2}). 


The photogate sensor feeds into  an \emph{Arduino nano board} \cite{WEB0}, which is programmed to count the number of slits passing through the sensor when the disk rotates as a result of the tube expansion. The Arduino sends a signal to a speaker that emits an audible beep per every count. In addition the Arduino stores the number of counts and, because it also stores the callibration of millimetres vs. counts, it can vocalize the corresponding millimeters of dilation. This vocalization is done utilizing a “Talkie library” software \cite{WEB1}, an Arduino Library developed by Peter Knight and Armin Joachimsmeyer, which consist of a software implementation of the Texas Instruments speech synthesis architecture from the late 1970s early 1980s. This vocalization is emitted when the S1 button is pressed (see Fig. \ref{fig4}(b)).


In our particular experience, the vocalization was only used as a verification  of the measurement, because the student usually uses the beeps as measurements. The synthetic voice is a help in the case that external noises impede the hearing of the beeps.  If one needs to repeat the measurement, button $S_{2}$ (see Fig. \ref{fig4}(b)) resets all stored numerical values, emitting two sound pulses. 
Additionally, a led (light-emitting diode) flashes every time a beep is emitted, for the teacher to verify the measurement.

The Arduino programming code implemented for this experience can be freely downloaded using the link provided in  Ref. \cite{WEB2}.


\section{Results and Discussion}

\begin{center}
\begin{table*}[t]
\begin{tabular}{ |c|c|c|c| }
\hline
\multicolumn{4}{|c|}{ 
Measurements taken by the student with impaired vision} \\
\hline
Measurement number (n) & $T$ $[^\circ$ C] & $\Delta L$ $[10^{-4}$ m] & $\Delta L/ L_{0}$ $[\times 10^{-4}$\\
\hline
 1 & 24.60  &  0.40 & 0.44\\
 2 &   32.20 & 1.20 & 1.31\\
 3 & 35.10 &  1.80 & 1.97\\
 4 & 37.60 &  2.40 & 2.62\\
 5 &  40.10 &     2.90 & 3.17\\
 6 & 42.50 &    3.50 & 3.83 \\
 7  & 44.80 &   4.10 & 4.48\\
 8 &  47.20 &   4.60 & 5.03\\
 9 & 49.90 &   5.20 & 5.68\\
 10 & 52.60 &  5.80 & 6.34\\
 11 & 54.90 &   6.40 & 7.00\\
 \hline
\end{tabular}
\caption{Measurements of the length expansion, $\Delta L$, as a function of temperature, $T$, of an Aluminum tube of room temperature length $L_{0}=0.915$ m. The length data were taken by the student with impaired vision and the temperature readings were registered by his laboratory partner.}
\label{table:1}
\end{table*}
\end{center}

In Table \ref{table:1} we show the data obtained by our study subject (SS) for the  thermal expansion of the aluminium tube. The temperature values $T$ were read by his partner (a student with normal vision) 
and the corresponding bar expansion values were read by SS.
Once SS finished collecting the data, he performed a linear regression where the slope indicates the value of the coefficient of linear expansion. 
In this particular lab experience, the value obtained was:

\begin{equation}
    \alpha_{Al}^{SS}\sim 22.86\times 10^{-6}\  \frac{1}{^\circ C}.
\end{equation}

\begin{figure}
    \centering
    \includegraphics[width= 1.0\columnwidth]{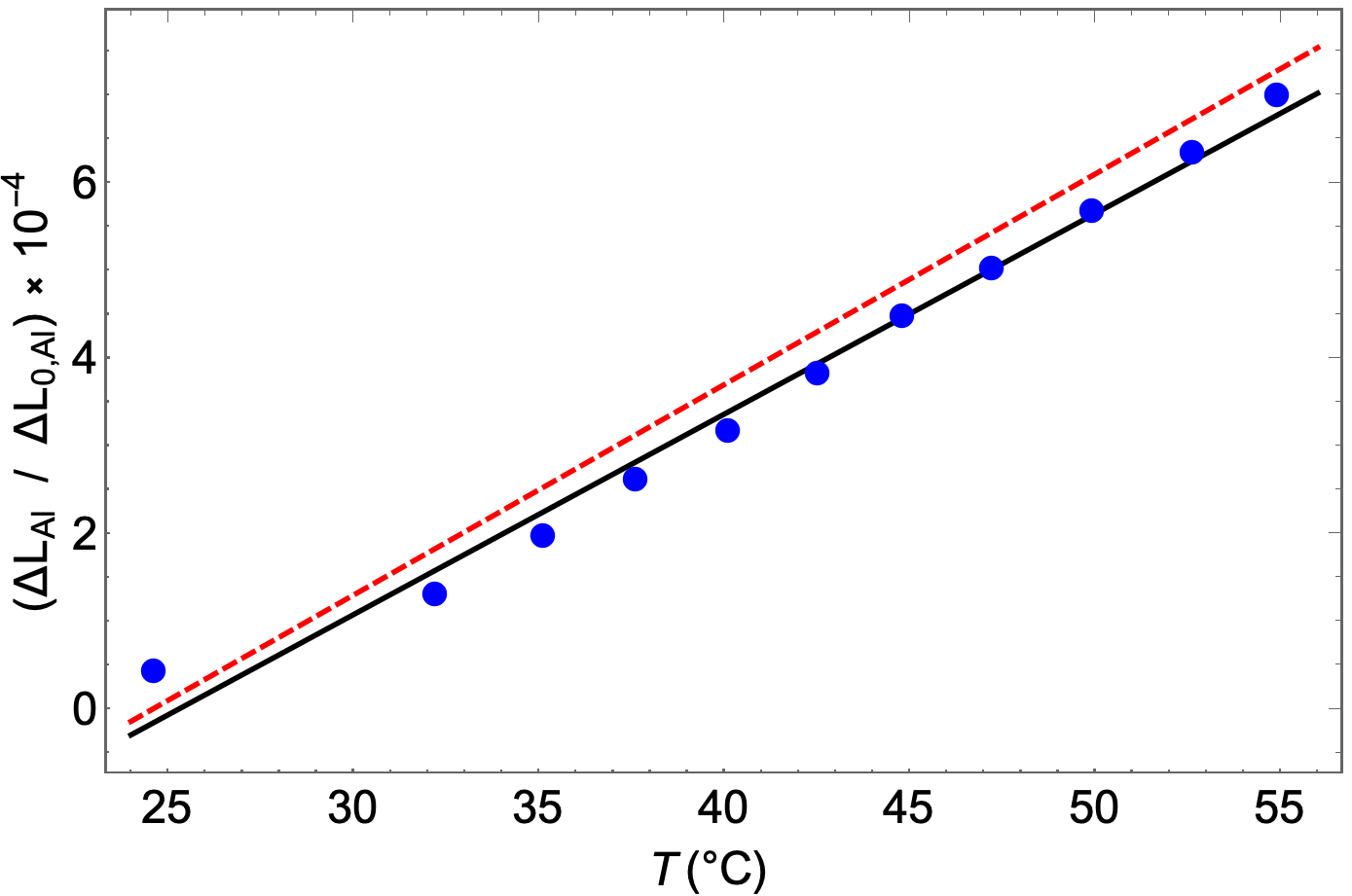}
    \caption{Data points from Table \ref{table:1} for the determination of the thermal expansion coefficient of the Aluminum tube. The solid line corresponds to the linear regression of the data and the dashed red line depicts the reference curve obtained with the listed value for Aluminum shown in Eq.~(\ref{eq2}).
    }
    \label{fig5}
\end{figure}

\begin{figure}
    \centering
    \includegraphics[width= 1.0\columnwidth]{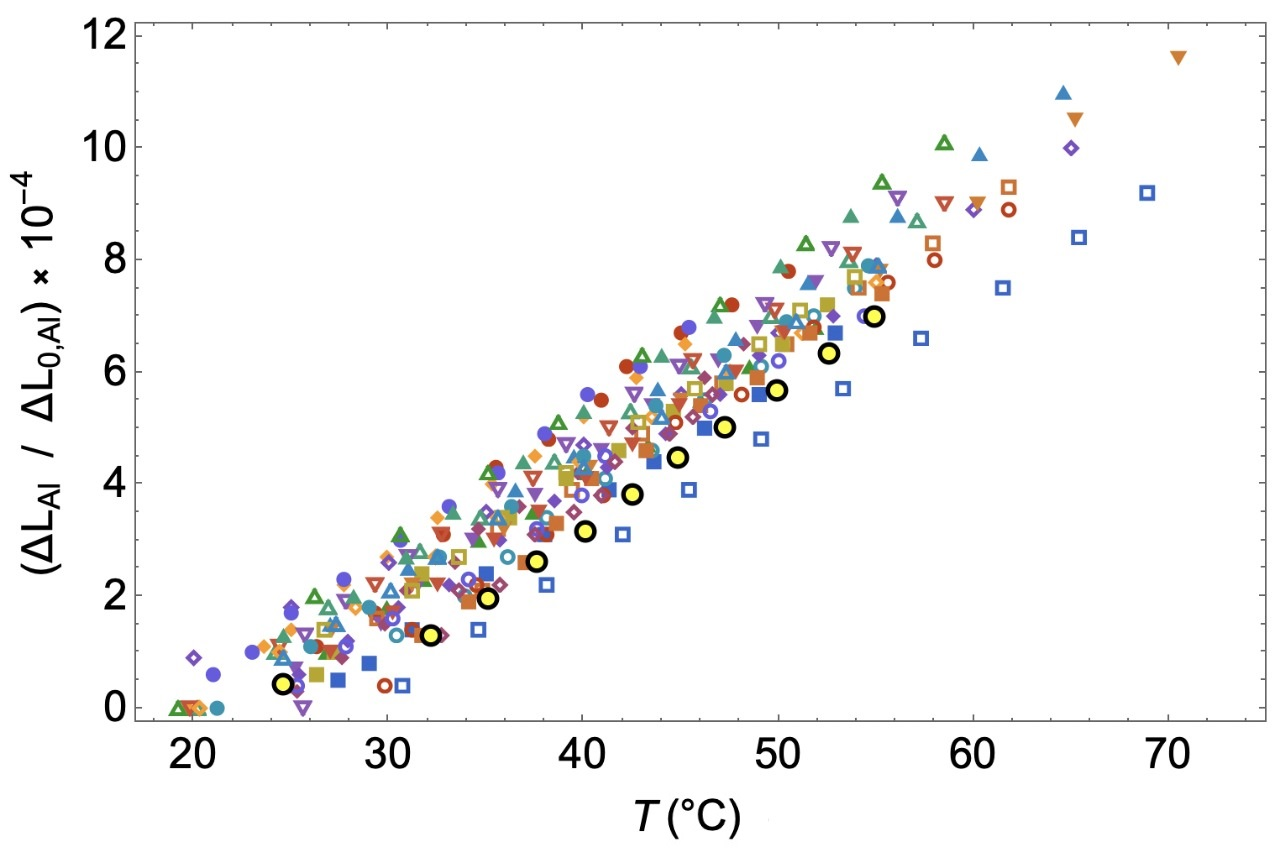}
    \caption{Data points of 29 reference student groups. The slope average is $\alpha_{Al}\sim 23.89\times 10^{-6}$ $[1/{^\circ C}]$. 
Yellow circles with black borders represent the data obtained by SS.}
    \label{fig6}
\end{figure}

The expected value for the Aluminium given by Eq. (\ref{eq2}) and the linear regression done by the student is shown in Fig. \ref{fig5}. 
This result is clearly within the expected range according to Eq. (\ref{eq2}).
To validate his result, we collected the results from other 29 group reports  (58 students), selected among the best grades of their respective classes (``high performance" groups). The data is presented in Fig. \ref{fig6}. 
The average value of the slope for all linear regressions made by these other groups is
\begin{equation}
\langle \alpha_{Al}^{HP}\rangle \sim 23.89\times 10^{-6}\  \frac{1}{^\circ C},
\end{equation}
which is also within the range of expected values given by Eq. (\ref{eq2}). 
The standard deviation of this set of values is
\begin{equation}
\label{standardeviation}
\Delta\alpha^{HP}_{Al}=4.81\times 10^{-6} \  \frac{1}{^\circ C},
\end{equation}
clearly showing that the value obtained by SS is well within the range of the measurements performed other by students. 

Nevertheless, we can still point out one particular source of systematic error of our experimental setup: the segmented disk allows for measurements of the tube dilatations in multiples of  0.01 mm only. This crucial point could be improved by using a smaller size photogate, or by enlarging the segmented disk; however the latter should not be pushed too far, because a larger disk implies a larger moment of inertia, which causes mechanical oscillations. 


\section{Conclusions}

We devised an adaptation for blind students of a teaching laboratory experience for the measurement of the thermal expansion coefficient of a metal tube. 
The purpose of our adaptation is to allow a blind student to fully participate in the measuring process, in addition to his/her usual participation in the data analysis and report.  Our adaptation consisted in modifying a length gauge usually in the shape of a dial, by connecting a segmented disk with a photogate sensor in replacement of the needle attached to the axis of the dial. In this way, the sensor reads the angle of rotation of the disk, which is proportional to the dilatation of the material (in this case, an Aluminum tube). The reading of the photogate sensor feeds into an Arduino platform which is programmed to count the number of slits passing through the photogate sensor and to send  signals to a speaker, converting in this way, a visual signal into one of the audible type.


We tested the device in a real laboratory session with a student who was blind from very young age. We found that the student was able to do the measurements with ease and his results were comparable to those of other student groups that had completed the experience successfully. We also found that the student felt much satisfaction by being actively integrated in the measurement process, in addition to his participation in the analysis and report, which was his usual duties in other experiences.



\section*{Acknowledgements}

F. J. P. acknowledges financial support from 
Conicyt (Chile) grant PAI77180015. P. V. and O.N. acknowledge  support from 
Conicyt (Chile) PIA/Basal FB 0807. O. N. acknowledge support from USM-DGIIP for Ph.D. fellowship. A. L. acknowledge support from PIA/Basal FB0821. C.D. was supported in part by Fondecyt (Chile) grant 1170171 and Conicyt (Chile) PIA/Basal FB0821.

\bibliographystyle{apsrev}

\begin{thebibliography}{27}
\expandafter\ifx\csname natexlab\endcsname\relax\def\natexlab#1{#1}\fi
\expandafter\ifx\csname bibnamefont\endcsname\relax
  \def\bibnamefont#1{#1}\fi
\expandafter\ifx\csname bibfnamefont\endcsname\relax
  \def\bibfnamefont#1{#1}\fi
\expandafter\ifx\csname citenamefont\endcsname\relax
  \def\citenamefont#1{#1}\fi
\expandafter\ifx\csname url\endcsname\relax
  \def\url#1{\texttt{#1}}\fi
\expandafter\ifx\csname urlprefix\endcsname\relax\def\urlprefix{URL }\fi
\providecommand{\bibinfo}[2]{#2}
\providecommand{\eprint}[2][]{\url{#2}}

\bibitem[{\citenamefont{Fermi}(1936)}]{Fermi_Thermo}
\bibinfo{author}{\bibfnamefont{E.}~\bibnamefont{Fermi}},
  \emph{\bibinfo{title}{Thermodynamics}} (\bibinfo{publisher}{Dover},
  \bibinfo{year}{1936}).

\bibitem[{\citenamefont{Callen}(1985)}]{Callen}
\bibinfo{author}{\bibfnamefont{H.~B.} \bibnamefont{Callen}},
  \emph{\bibinfo{title}{Thermodynamics and an Introduction to
  Thermostatistics}} (\bibinfo{publisher}{John Wiley and Sons},
  \bibinfo{year}{1985}), \bibinfo{edition}{2nd} ed.

  
\bibitem[{\citenamefont{Holt}(2017)}]{Holt}
\bibinfo{author}{\bibfnamefont{M.} \bibnamefont{Holt}},
  \bibinfo{author}{\bibfnamefont{D.}~\bibnamefont{Gillen}}, 
  \bibinfo{author}{\bibfnamefont{S.~D.} \bibnamefont{Nandlall}},
  \bibinfo{author}{\bibfnamefont{K.} \bibnamefont{Setter}},
    \bibinfo{author}{\bibfnamefont{P.} \bibnamefont{Thoman}} 
    \bibinfo{author}{\bibfnamefont{S.~A.} \bibnamefont{Kane}},
    \bibinfo{author}{\bibfnamefont{C.~H.} \bibnamefont{Miller}},
    \bibinfo{author}{\bibfnamefont{C.} \bibnamefont{Cook}} \bibnamefont{and}
    \bibinfo{author}{\bibfnamefont{C.} \bibnamefont{Supalo}},
  \bibinfo{journal}{Phys. Teach.} \textbf{\bibinfo{volume}{57}},
  \bibinfo{pages}{94} (\bibinfo{year}{2019}).
  
    
   \bibitem[{\citenamefont{DeBuvitz}(2019)}]{DeBuvitz}
\bibinfo{author}{\bibfnamefont{W.} \bibnamefont{DeBuvitz}},
  \bibinfo{journal}{The Phys. Teach.} \textbf{\bibinfo{volume}{57}},
  \bibinfo{pages}{276} (\bibinfo{year}{2019}).
  
  \bibitem[{\citenamefont{Marinho et~al.}(2016)}]{Marinho}
\bibinfo{author}{\bibfnamefont{F.} \bibnamefont{Marinho}}, \bibnamefont{and}
  \bibinfo{author}{\bibfnamefont{L.} \bibnamefont{Paulucci}},
  \bibinfo{journal}{Eur. J. Phys.} \textbf{\bibinfo{volume}{37}},
  \bibinfo{pages}{025003} (\bibinfo{year}{2016}).

  

\bibitem[{\citenamefont{Goncalves et~al.}(2017)}]{Goncalves1}
\bibinfo{author}{\bibfnamefont{A.~M.~B.} \bibnamefont{Goncalves}},
  \bibinfo{author}{\bibfnamefont{C.~R.} \bibnamefont{Cena}}, \bibnamefont{and}
  \bibinfo{author}{\bibfnamefont{D.~F.} \bibnamefont{Bozano}},
  \bibinfo{journal}{Phys. Educ.} \textbf{\bibinfo{volume}{52}},
  \bibinfo{pages}{043002} (\bibinfo{year}{2017}).

\bibitem[{\citenamefont{Keith Atkin.}(2016)}]{Atkin}
\bibinfo{author}{\bibfnamefont{K.}\bibnamefont{Atkin}},\bibinfo{journal}{Phys.
  Educ.} \textbf{\bibinfo{volume}{51}}, \bibinfo{pages}{065006}
  (\bibinfo{year}{2016}).

\bibitem[{\citenamefont{KuvNob}(2015)}]{KuvNob}
\bibinfo{author}{\bibfnamefont{S.}~\bibnamefont{Kub{\'{i}}}nov{\'{a}}} \bibnamefont{and}
  \bibinfo{author}{\bibfnamefont{J.}~\bibnamefont{ {\v{S}}l{\'{e}}gr}}, \bibinfo{journal}{Phys.
  Educ.} \textbf{\bibinfo{volume}{50}}, \bibinfo{pages}{472}
  (\bibinfo{year}{2015}).

\bibitem[{\citenamefont{Goncalves2}(2017)}]{Goncalves2}
\bibinfo{author}{\bibfnamefont{A.~M.~B.} \bibnamefont{Goncalves}},
  \bibinfo{author}{\bibfnamefont{C.~R.}\bibnamefont{Cena}}, 
  \bibinfo{author}{\bibfnamefont{D.~C.~D.} \bibnamefont{Alves}},
  \bibinfo{author}{\bibfnamefont{N.~C.~G.} \bibnamefont{Errobidart}},
    \bibinfo{author}{\bibfnamefont{M.~I.~A.} \bibnamefont{Jardim}} \bibnamefont{and}
    \bibinfo{author}{\bibfnamefont{W.~P.} \bibnamefont{Queiros}}.
  \bibinfo{journal}{Phys. Educ.} \textbf{\bibinfo{volume}{52}},
  \bibinfo{pages}{053002} (\bibinfo{year}{2017}).
  
  \bibitem[{\citenamefont{Toenders}(2017)}]{Toenders}
\bibinfo{author}{\bibfnamefont{F.~G.~C.} \bibnamefont{Toenders}},
  \bibinfo{author}{\bibfnamefont{L.~G.~A.}\bibnamefont{Putter-Smits}}, 
  \bibinfo{author}{\bibfnamefont{W.~T.~M.} \bibnamefont{Sanders}} \bibnamefont{and}
    \bibinfo{author}{\bibfnamefont{P.} \bibnamefont{den Brok}}.
  \bibinfo{journal}{Phys. Educ.} \textbf{\bibinfo{volume}{52}},
  \bibinfo{pages}{055006} (\bibinfo{year}{2017}).


  \bibitem[{\citenamefont{Galeriu1}(2013)}]{Galeriu1}
\bibinfo{author}{\bibfnamefont{C.} \bibnamefont{Galeriu}},
  \bibinfo{journal}{Phys. Teach.} \textbf{\bibinfo{volume}{51}},
  \bibinfo{pages}{156} (\bibinfo{year}{2013}).


  \bibitem[{\citenamefont{Galeriu2}(2014)}]{Galeriu2}
\bibinfo{author}{\bibfnamefont{C.} \bibnamefont{Galeriu}},
\bibinfo{author}{\bibfnamefont{S.} \bibnamefont{Edwards}} and
\bibinfo{author}{\bibfnamefont{G.} \bibnamefont{Esper}},
  \bibinfo{journal}{Phys. Teach.} \textbf{\bibinfo{volume}{52}},
  \bibinfo{pages}{157} (\bibinfo{year}{2014}).
  
  
   \bibitem[{\citenamefont{Galeriu3}(2018)}]{Galeriu3}
\bibinfo{author}{\bibfnamefont{C.} \bibnamefont{Galeriu}},
  \bibinfo{journal}{Phys. Teach.} \textbf{\bibinfo{volume}{56}},
  \bibinfo{pages}{618} (\bibinfo{year}{2018}).
  
  
    \bibitem[{\citenamefont{Zou}(2011)}]{Zou}
\bibinfo{author}{\bibfnamefont{J. J.}\bibnamefont{Millis}}.
 \bibinfo{title}{Blind Florida State U. students sue over e-learning systems, The Chronicle of Higher Education, 2011.}
  \bibinfo{journal}{Retrieved from https://www.chronicle.com/blogs/ wiredcampus/blind-florida-state-u-students-sue-over- e-learning-systems. [July. 2019]}.
  
     \bibitem[{\citenamefont{Lewin}(205)}]{Lewin}
\bibinfo{author}{\bibfnamefont{J.}~\bibnamefont{Lewin}}.
 \bibinfo{title}{Havard and M.I.T. are Sued Over Lack of Closed Captions, The New York Times, 2015. }
  \bibinfo{journal}{Retrieved from https://www.nytimes.com/2015/02/13/education/harvard- and-mit-sued-over-failing-to-caption-online-courses.html.  [July. 2019]}.
  
  
  \bibitem[{\citenamefont{Scanlon}(2018)}]{Scanlon}
\bibinfo{author}{\bibfnamefont{E.}\bibnamefont{Scanlon}},
  \bibinfo{author}{\bibfnamefont{J.}\bibnamefont{Schreffler}}, 
  \bibinfo{author}{\bibfnamefont{W.}\bibnamefont{James}},
  \bibinfo{author}{\bibfnamefont{E.} \bibnamefont{Vasquez}} \bibnamefont{and}
    \bibinfo{author}{\bibfnamefont{J.~J.} \bibnamefont{Chin}}.
  \bibinfo{journal}{Phys. Rev. Phys. Educ. Res.} \textbf{\bibinfo{volume}{14}},
  \bibinfo{pages}{020102} (\bibinfo{year}{2018}).
  
   
 
  
        \bibitem[{\citenamefont{Millis}(1996)}]{Millis}
\bibinfo{author}{\bibfnamefont{B. J.}~\bibnamefont{Millis}}.
 \bibinfo{title}{Materials presented at The University of Tennessee at Chattanooga Instructional Excellence Retreat, 1996.}
  \bibinfo{journal}{ Retrieved from https://www.utc.edu/walker-center-teaching-learning/teaching-resources/cooperative-learning.php. [July. 2019]}.
  
       \bibitem[{\citenamefont{WEB3}(2019)}]{WEB3}
\bibinfo{author} {\bibnamefont{Dea.usm.cl/2017/07/26/modelo-educativo. Educational model of the Technical University Federico Santa Maria (Chile) [Online]. 2019.}}
  \bibinfo{journal}{http://dea.usm.cl/2017/07/26/modelo-educativo/ [July. 2019]}.
  

           \bibitem[{\citenamefont{software1}(2019)}]{software1}
\bibinfo{author} {\bibnamefont{Dti.usm.cl/servicios6/. Office 365 A1 Plus for faculty [Online]. 2019.}}
  \bibinfo{journal}{http://www.dti.usm.cl/servicios6/ [July. 2019]}.
  
  \bibitem[{\citenamefont{WEB00}(2020)}]{WEB00}
\bibinfo{author} {\bibnamefont{Engineeringtoolbox.com/linear-expansion-coefficients-d\textunderscore 95.html. Engineering toolbox. Linear temperature expansion coefficients for aluminum, copper, glass, iron and other common materials [Online]. 2020.}}
  \bibinfo{journal}{  https://engineeringtoolbox.com/linear-expansion-coefficients-d\textunderscore 95.html  [Jan. 2020]}.  
    
  
   \bibitem[{\citenamefont{Fermi}(2008)}]{Freedman}
\bibinfo{author}{\bibfnamefont{H. D.}\bibnamefont{Young}} and
\bibinfo{author}{\bibfnamefont{R. A.} \bibnamefont{Freedman}},
  \emph{\bibinfo{title}{University Physics with Modern Physics}} (\bibinfo{publisher}{Pearson Education},
  \bibinfo{year}{2008, $12^{th}$ ed.})

  
  
     \bibitem[{\citenamefont{WEB0}(2020)}]{WEB0}
\bibinfo{author} {\bibnamefont{Arduino.cc/en/guide/introduction. Introduction to Arduino [Online]. 2020.}}
  \bibinfo{journal}{https://arduino.cc/en/guide/introduction [Jan. 2020]}.  
  
     \bibitem[{\citenamefont{WEB1}(2019)}]{WEB1}
\bibinfo{author} {\bibnamefont{Github.com/going-digital/Talkie. Texas Instruments speech synthesis architecture (Linear Predictive Coding) [Online]. 2019.}}
  \bibinfo{journal}{https://github.com/going-digital/Talkie [May. 2019]}.
  

     \bibitem[{\citenamefont{WEB2}(2019)}]{WEB2}
\bibinfo{author} {\bibnamefont{Github.com/PHYS-USM/lab\textunderscore code. USM laboratory code for a linear expansion experience [Online]. 2019.}}
  \bibinfo{journal}{https://github.com/PHYS-USM/lab\textunderscore code [Jun. 2019]}.
  
\end{thebibliography}

\end{document}